\documentclass[%
prl
 aip,
 jmp,%
 amsmath,amssymb,
twocolumn,superscriptaddress]{revtex4-2}

\usepackage{graphicx}
\usepackage{dcolumn}
\usepackage{bm}
\usepackage{xcolor}
\usepackage{booktabs}

\newcommand{\NIST}{Sensor Science Division, National Institute of Standards and Technology, Gaithersburg, Maryland 20899, USA}

\newcommand{\PNPI}{Petersburg Nuclear Physics Institute of NRC ``Kurchatov Institute'', Gatchina, Leningrad district 188300, Russia}  

\newcommand{\UD}{University of Delaware, Newark, Delaware, USA}

\begin{document}

\title{Laser Spectroscopy of the y$^7$P$_J^{\circ}$ states of Cr I}

\date{\today}

\author{E. B. Norrgard}

 \affiliation{\NIST}
 
\author{D. S. Barker}

 \affiliation{\NIST}
 
 \author{S. P. Eckel}
 \affiliation{\NIST}
 
 \author{S. G. Porsev}
  \affiliation{\UD}   \affiliation{\PNPI} 

 \author{C. Cheung}
  \affiliation{\UD} 
 
 \author{M. G. Kozlov}
\affiliation{\PNPI}
\affiliation{St.~Petersburg Electrotechnical University
``LETI'', Prof. Popov Str. 5, St.~Petersburg, 197376, Russia}
 
\author{I. I. Tupitsyn}
\affiliation{Department of Physics, St. Petersburg State University,
Ulianovskaya 1, Petrodvorets, St.Petersburg, 198504, Russia}

\author{M. S. Safronova}
 \affiliation{\UD}

\begin{abstract}
Here we report measured and calculated values of decay rates of the 3d$^4$($^5$D)4s4p($^3$P$^{\rm{o}}$)\ y$^7$P$^{\rm{o}}_{2,3,4}$ states  of Cr I. 
The decay rates are measured using time-correlated single photon counting  with roughly 1\,\% total uncertainty. In addition, 
the isotope shifts for these transitions are measured by laser induced fluorescence to roughly 0.5\,\% uncertainty.  The decay rate calculations are carried out by a hybrid approach that combines configuration interaction and the linearized coupled cluster method (CI+all-order method). The measurements provide a much needed precision benchmark for testing the accuracy of the CI+all-order approach for such  complicated systems with six valence electrons, allowing to significantly expand its applicability. These measurements also demonstrate operation of a cryogenic buffer gas beam source for future experiments with MgF molecules toward quantum blackbody thermometry.

\end{abstract}

\maketitle

\section{Introduction}
The decay rates of excited quantum states are important parameters for characterizing the electronic structure of atomic and molecular systems.  Precise decay rate measurements are necessary for atomic studies of electroweak symmetries \cite{Porsev2006,Aubin2004,Toh2019}, calibration of astrophysical measurements \cite{Marek1975, Becker1977}, prediction of optical trap depths and magic wavelengths \cite{safronova2016,Griesmaier2007},  fluorescence measurements of the number of magneto-optically trapped atoms or molecules \cite{Ivory2014,Barry2014}, calibrating atom-based sensors \cite{Norrgard2021}, and more.

Here we report measured and calculated values of the $1/e$ decay rates of the 3d$^4$($^5$D)4s4p($^3$P$^\circ$)\ y$^7$P$^\circ_{2,3,4}$ states (hereafter, y$^7$P$^\circ_J$ states) of Cr I.  We also report the isotope shifts of these transitions from the ground state.  The isotope shifts are measured by laser induced fluorescence to roughly 0.5\,\% uncertainty, while the decay rates are measured using time-correlated single photon counting (TCSPC) with roughly 1\,\% total uncertainty. TCSPC techniques have previously  been used to measured the decay rates of several atoms \cite{Young1994,Aubin2004,Gomez2004,Toh2019} and molecules \cite{Aggarwal2019} to 1\,\% uncertainty or better.

We had two primary motivations for carrying out these measurements.  First, these measurements demonstrate operation of a cryogenic buffer gas beam source constructed at NIST to create MgF molecules for laser cooling and trapping experiments toward quantum blackbody thermometry \cite{Norrgard2021}.  In particular, the  transition from the ground 3d$^5$($^6$S)4s\ a$^7$S$_3$ state of Cr to the y$^7$P$^{\rm{o}}_3$  state has nearly the same frequency ($\nu = 834.0286$\,THz \cite{NIST_ASD}) as the $X^2\Sigma^+, v=0, N=1 \rightarrow A^2\Pi_{1/2}, v=0, J^P=1/2^+$ main cycling transition of MgF ($\nu =834.2903$\,THz \cite{Xu2019}).  The decay rates of the Cr y$^7$P$^{\rm{o}}_J$ states \cite{Marek1973,Becker1977,Cooper1997,NIST_ASD} are also within 10\,\% of the expected decay rate for the MgF cycling transition \cite{Kang2015}.  Cr thus serves are a suitable proxy for commissioning the cryogenic buffer gas beam source, lasers, and diagnostic systems for future work with MgF.

Second, these measurements provide a high-precision benchmark for the first application of the  configuration interaction (CI) and the linearized coupled cluster method (CI+all-order method) for a system with 6 valence electrons.  The CI+all-order method was demonstrated to produce accurate results for a wide range of atomic properties for atoms and ions with 2 to 4 valence electrons \cite{CheSafPor21,ZhaStuWeb20,YamSafGib19,SafPorKoz18}. This method has provided precision data for many applications, including development of ultra-precise clocks \cite{CheSafPor21,YamSafGib19,Safronova2013},  studies of fundamental symmetries \cite{PorSafKoz12,PorKozSaf16}, searches for the variation of fundamental constants \cite{SafDzuFla14}, quantum simulation and computation \cite{ZhaBisBro14,HeiParSan20}, plasma physics \cite{SafSafNak17}, nuclear physics \cite{SafPorKoz18,Rea18}, and many others. It was  previously thought to be unfeasible to apply this method to systems such as Cr due to extremely large number of configurations that have to be considered as well as limitation of the initial closed-shell approximation. In this work, we demonstrated a 1\,\% to 2.5\,\% agreement with experiment for Cr electric-dipole transition matrix elements and a few percent accuracy for a wide range of energy levels.  We also tested the intrinsic accuracy of the approach by demonstrating the ability to saturate a CI space in such computation to a good numerical precision.  The methodology used here is not specific to Cr and we expect to be able to apply CI+all-order method  for a wide range of other systems with 5 to 6 valence electrons, which opens many future applications. 

To our knowledge, these are the highest precision decay rates reported to date for any Cr I levels.
 The decay rates of the y$^7$P$^{\rm{o}}_J$ were previously measured  by Becker \textit{et al.}  \cite{Becker1977} with roughly 3\,\% uncertainty using laser excitation; by Cooper \textit{et al.} \cite{Cooper1997} with 4.5\,\% uncertainty using laser excitation; and by Marek and Richter \cite{Marek1973} to roughly 10\,\% uncertainty by the phase shift method. 
Tian \textit{et al.} \cite{Tian2015} have measured decay rates of 43 Cr I levels with an uncertainty  of a few percent using pulsed laser excitation in the wavelength range 216\,nm to 649\,nm.  Despite this extensive survey, the y$^7$P$^{\rm{o}}_J$ states were omitted, apparently due to a gap in the coverage of their dye laser. Another compilation of Cr I decay rates is found in Tozzi \textit{et al.}  \cite{Tozzi1985}, which covers lines  in the 290\,nm to 900\,nm range to roughly 7\,\% accuracy by Fourier transform spectroscopy on a chromium hollow cathode lamp.  
Pulsed laser excitation methods have been used to measure the decay rates of the z$^7$P$^{\rm{o}}_J$ levels to roughly 3\,\% precision by Measures \textit{et al.} \cite{Measures1977}, and to roughly 2\,\% precision for z$^7$P$^{\rm{o}}_4$ only by Hannaford and Lowe \cite{hannaford1981}. Marek \cite{Marek1975} also measured the decay rates of nine Cr I levels in the 425\,nm to 429\,nm range including z$^7$P$^{\rm{o}}_J$ to several percent precision.  Isotope shifts of several other low-lying Cr I transitions were reported in Frumann \textit{et al.} \cite{Furmann2005}.

\section{Theory}
\begin{table} [th]
\caption{\label{tab1t} Comparison of theoretical energy levels (in cm$^{-1}$) with experiment \cite{NIST_ASD}. }
\begin{ruledtabular}
\begin{tabular}{lcrrrr}
\multicolumn{1}{c}{Level}& \multicolumn{1}{c}{Term}&  \multicolumn{1}{c}{Expt.}&
  \multicolumn{1}{c}{Theory}&\multicolumn{1}{c}{Diff.}& \multicolumn{1}{c}{Diff. (\%)}\\
\hline
$   J=3$ even  &             &           &          &         &           \\
$   3d^5 4s   $&$   \rm{a}^7\rm{S}   $   &   0       &   0      &         &      \\
$   3d^4 4s^2 $&$   \rm{a}^5\rm{D}   $   &   8095    &   7128   &   967   &   12\%   \\
$   3d^5 4s   $&$   \rm{a}^5\rm{G}   $   &   20521   &   21283  &   -762  &   -3.7\%   \\
$   3d^5 4s   $&$   \rm{a}^5\rm{P}   $   &   21841   &   22702  &   -861  &   -3.9\%   \\ [0.5pc]
$   J=2$ even&             &           &          &         &           \\
$   3d^5 4s   $&$  \rm{a}^5\rm{S}  $   &   7593    &   8040   &   -447  &   -5.9\%   \\
$   3d^4 4s^2 $&$  \rm{a}^5\rm{D}  $   &   7927    &   7171   &   756   &   9.5\%   \\
$   3d^5 4s   $&$  \rm{a}^5\rm{G}  $   &   20517   &   21386  &   -869  &   -4.2\%   \\
$   3d^5 4s   $&$  \rm{a}^5\rm{P}  $   &   21848   &   22857  &   -1010 &   -4.6\%   \\
$   3d^4 4s^2 $&$  \rm{a}^3\rm{P}  $   &   24093   &   24109  &   -15   &   -0.1\%   \\
$   3d^5 4s   $&$  \rm{b}^5\rm{D}  $   &   24300   &   25087  &   -787  &   -3.2\%   \\     [0.5pc]
$   J=4$ odd  &             &           &           &         &           \\
$   3d^5 4p   $&$  \rm{z}^7\rm{P}^{\circ}  $   &   23499   &   23596   &   -97   &   -0.4\%   \\
$   3d^4 4s4p $&$  \rm{z}^7\rm{F}^{\circ}  $   &   25360   &   24545   &   815   &   3.2\%   \\
$   3d^4 4s4p $&$  \rm{z}^7\rm{D}^{\circ}  $   &   27650   &   27045   &   605   &   2.2\%   \\
$   \bm{3d^4 4s4p} $&$  \bf{y^7P}^{\circ}  $   &   \bf{27935}   &   \bf{28629}   &   \bf{-694}  &   \bf{-2.5\%}   \\
$   3d^4 4s4p $&$  \rm{z}^5\rm{F}^{\circ}  $   &   31106   &   30588   &   518   &   1.7\%   \\    [0.5pc]
$   J=3$ odd &            &           &           &         &      \\
$   3d^5 4p   $&$  \rm{z}^7\rm{P}^{\circ}  $   &   23386   &   23482   &   -96   &   -0.4\%   \\
$   3d^4 4s4p $&$  \rm{z}^7\rm{F}^{\circ}  $   &   25206   &   24395   &   811   &   3.2\%   \\
$   3d^5 4p   $&$  \rm{z}^5\rm{P}^{\circ}  $   &   26787   &   28099   &   -1312 &   -4.9\%   \\
$   3d^4 4s4p $&$  \rm{z}^7\rm{D}^{\circ}  $   &   27500   &   26900   &   600   &   2.2\%   \\
$   \bm{3d^4 4s4p} $&$  \bf{y^7P}^{\circ}  $   &   \bf{27820}   &   \bf{28528}   &   \bf{-707}  &  \bf{ -2.5\%}   \\
$   3d^4 4s4p $&$  \rm{y}^5\rm{P}^{\circ}   $   &   29825   &   29114   &   711   &   2.4\%   \\
$   3d^4 4s4p $&$  \rm{y}^5\rm{F}^{\circ}   $   &   30965   &   30448   &   517   &   1.7\%   \\
$   3d^4 4s4p $&$  \rm{y}^5\rm{D}^{\circ}   $   &   33672   &   33412   &   259   &   0.8\%   \\
$   3d^4 4s4p $&$  \rm{z}^3\rm{F}^{\circ}   $   &   36034   &   35877   &   157   &   0.4\%   \\
$   3d^4 4s4p $&$  \rm{z}^3\rm{D}^{\circ}   $   &   38911   &   38883   &   29   &   0.1\%   \\   [0.5pc]
$   J=2$ odd &           &           &           &         &      \\
$   3d^5 4p   $&$ \rm{z}^7\rm{P}^{\circ}   $   &   23305   &   23392   &   -87   &   -0.4\%   \\
$   3d^4 4s4p $&$ \rm{z}^7\rm{F}^{\circ}   $   &   25089   &   24282   &   808   &   3.2\%   \\
$   3d^5 4p   $&$ \rm{z}^5\rm{P}^{\circ}   $   &   26796   &   28079   &   -1282   &   -4.8\%   \\
$   3d^4 4s4p $&$ \rm{z}^7\rm{D}^{\circ}   $   &   27382   &   26785   &   597   &   2.2\%   \\
$   \bm{3d^4 4s4p} $&$ \bf{y^7P}^{\circ}   $   &   \bf{27729}   &   \bf{28441}   &   \bf{-712}   &   \bf{-2.6\%}   \\
$   3d^4 4s4p $&$ \rm{y}^5\rm{P}^{\circ}    $   &   29585   &   28886   &   698   &   2.4\%   \\
$   3d^4 4s4p $&$ \rm{z}^5\rm{F}^{\circ}    $   &   30859   &   30343   &   515   &   1.7\%   \\
$   3d^4 4s4p $&$ \rm{z}^5\rm{D}^{\circ}    $   &   33542   &   33282   &   260   &   0.8\%   \\
$   3d^4 4s4p $&$ \rm{z}^3\rm{P}^{\circ}    $   &   34190   &   33757   &   433   &   1.3\%   \\
$   3d^4 4s4p $&$ \rm{z}^3\rm{F}^{\circ}    $   &   35898   &   35735   &   163   &   0.5\%   \\
$   3d^4 4s4p $&$ \rm{z}^3\rm{D}^{\circ}    $   &   38731   &   38704   &   26   &   0.1\%   \\
  \end{tabular}
\end{ruledtabular} \end{table}

We calculated Cr energies and transition rates using the method that combines linearized coupled-cluster and configuration interaction approaches; the CI+all-order  method \cite{SafKozJoh09}.  In this method, the coupled-cluster approach is used to construct an effective Hamiltonian $H_{\rm{eff}}$ that includes core and core-valence correlations. The many-electron wave function is obtained using the CI method as a linear combination of all distinct many-electron states of a given angular momentum $J$ and parity:
$$ \Psi_{J} = \sum_{i} c_{i} \Phi_i.$$
The energies and wave functions of the low-lying states are determined by diagonalizing this effective rather than bare Hamiltonian.

The CI+all-order method was demonstrated to produce accurate results for a wide range of atomic properties for atoms and ions with 2 to 4 valence electrons \cite{CheSafPor21,ZhaStuWeb20,YamSafGib19,SafPorKoz18}. The main problem in extending this approach to more complicated systems such as Cr is an exponential increase in the number of configurations in the CI expansion with the number of valence electrons.  Another problem is the degrading quality of one-electron functions, which are built in a closed-shell Dirac-Hartree-Fock potential, i.e., $V^{N-M}$, where $N$ is the total number of electrons and $M$ is the number of valence electrons.

In some cases a different potential can be used to construct one-electron orbitals for the CI+all-order calculations. For example, in Ref.~\cite{PorKozSaf16}  Pb I
was considered as a system with four valence
electrons but the $V^{N-2}$ potential was used for the construction of
the basis-set orbitals. Such an approach is expected to provide better initial 
approximation but it is more complicated 
and the area of its applicability requires further research. 
While one can construct one-electron orbitals for partially open shells, an all-order effective Hamiltonian $H_{\rm{eff}}$ cannot be constructed for such a basis.

The problems mentioned above hampered application of the effective Hamiltonian approach to the systems with more than 4 valence electrons. However, in 2005 \citet{Dzu05a} used this method to calculate properties of the Kr atom. An effective Hamiltonian was formed using second order many-body perturbation theory in the $V^{N-8}$ potential and Kr was considered as an 8-electron system. Only the ground state was calculated and 2.4\,\% accuracy for the ionization potential was obtained. 

In this work, we have successfully applied the CI+all-order approach for the case of six valence electrons for the first time. We start from  Dirac-Hartree-Fock one-electron wave functions for the low-lying valence electrons: $3d$, $4s$, $4p$, $5s$, $4d$, $5p$, and $4f$ with the $1s^22s^22p^63s^23p^6$ core, i.e., constructed in a $V^{N-6}$ potential.
 All other orbitals (up to $35spdfghi$) are constructed in a spherical cavity using B-splines. Such a finite basis set method discretizes the continuum spectrum: a sum over the finite basis is equivalent (to a numerical precision) to the sum over all bound states and integration over the continuum. This large basis is used for all coupled-cluster computations but can be reduced for CI computations. Taking into account that we are starting from a potential of the Cr$^{6+}$ core, which is significantly more compact than neutral Cr, we reduced the cavity size from 60~$a_0$ used for 2 to 3 valence electrons to 30~$a_0$ ($a_0\,\approx\,52.9$\,pm is the Bohr radius).
 
\begin{table*} [th]
\caption{\label{tab2t} Electric-dipole reduced matrix elements (in $ea_0$, where $e$ is the electron charge and $a_0$ is the Bohr radius) and transition rates (in s$^{-1}$) in Cr.
The CI+all-order values include RPA correction; the contributions from the core-Brueckner ($\sigma$), two-particle
(2P), structural radiation (SR), and normalization (Norm) corrections are listed separately.
 Final matrix elements are given in column``Total''. The transition rates are given in the last column. }
\begin{ruledtabular}
\begin{tabular}{lccccccc}
\multicolumn{1}{c}{Transition}& \multicolumn{1}{c}{CI+all-order+RPA}&  \multicolumn{1}{c}{$\sigma$}&
  \multicolumn{1}{c}{SR}&\multicolumn{1}{c}{2P}& \multicolumn{1}{c}{Norm} & \multicolumn{1}{c}{Total}
  & \multicolumn{1}{c}{Tr. rate} \\
\hline	
$3d^44s4p$~ $\rm{y}^7\rm{P}_4^{\circ}-3d^54s$ $\rm{a}^7\rm{S}_3$&	6.089&	0.004&	0.010&	0.038&	-0.704&	5.437&	145$\times10^6$\,s$^{-1}$  \\
$3d^44s4p$~ $\rm{y}^7\rm{P}_3^{\circ}-3d^54s$ $\rm{a}^7\rm{S}_3$&	5.295&	0.003&	0.008&	0.027&	-0.611&	4.722&	139$\times10^6$\,s$^{-1}$  \\
$3d^44s4p$~ $\rm{y}^7\rm{P}_2^{\circ}-3d^54s$ $\rm{a}^7\rm{S}_3$&	4.476&	0.002&	0.007&	0.022&	-0.511&	3.996&	138$\times10^6$\,s$^{-1}$  \\
  \end{tabular}
\end{ruledtabular} \end{table*}

Usually, we construct the set of CI configurations  by starting from the configurations of interest, here $3d^54s$, $3d^44s^2$, $3d^54p$, and $3d^44s4p$, and make all possible single and double excitations to a large basis, such as $20spd18f16g$. Then, one accounts for some triple and quadruple excitations, which are generally much smaller.
However, because the one-electron orbitals were built in the $V^{N-6}$ potential, $4s$ orbital for neutral Cr is a mixture of our $4s$, $5s$, and higher $ns$ basis set orbitals. This leads to increased weights of the $3d^55s$, $3d^44d4s$, $3d^44s5p$, etc. configurations. A solution is to use all such configurations as basis configurations and allow all single and double excitations for such configurations into as many basis set states as numerically feasible. In addition, we allow all single excitations from a large number of dominant configurations ($\sim$130 non-relativistic configurations) to the $20spd18f16g$ basis when constructing the CI space for odd states.

An obvious problem is a very large number of resulting configurations and corresponding Slater determinants (207,000 and 223,000 relativistic configurations and 45 million and 51 million Slater determinants for odd and even states, respectively). We note that even and odd states are computed separately in CI. Our recently developed Message Passing Interface (MPI) version of the CI code \cite{CheSafPor20,CheSafPor21} allows us to carry our computations with such a large number of configurations in a reasonable time, using 500 to 1000 central processing units (CPUs). Using very large number of  configurations also significantly affects convergence of the Davidson procedure that extracts low-lying eigenvalues and eigenstates.
We carried out additional improvements to the code, allowing for fast diagonalization of a much larger matrix used to initialize a Davidson procedure \cite{Dav75}, improving convergence. To further alleviate the convergence issues, we carry out computations for different $J$  separately, reordering configurations by weight and omitting those with negligible weights, reducing the size of the computations.

 The results for the energy levels in cm$^{-1}$ are listed in Table~\ref{tab1t}, grouped by $J$ and parity. Comparing with experimental energy values \cite{NIST_ASD} required resolving another issue, i.e. correct identification of the $\rm{y}^7\rm{P}^{\circ}$ levels, as $\rm{z}^7\rm{D}^{\circ}$ levels have similar energy and both $^5\rm{P}^{\circ}$ levels mix with the $\rm{y}^7\rm{P}^{\circ}$ levels for $J=2$ and $J=3$ states (there is no
 $^5\rm{P}^{\circ}$ level with $J=4$).
We wrote a code that computes the expectation values of $\langle\bm{L}^2\rangle\ $ and
$\langle\bm{S}^2\rangle\ $with the final wave functions, where $\bm{L}$ and $\bm{S}$ are total electron orbital angular momentum and spin operators, respectively.
Computing these quantities for all states allowed us to find approximate quantum numbers $L$ and $S$, where $\langle\bm{L}^2\rangle\ =L(L+1)$ and $\langle\bm{S}^2\rangle\ =S(S+1)$, and unambiguously identify all terms in Table~\ref{tab1t}. The energies are generally in good agreement with the experiment. The differences between the theory and experiment for $\rm{y}^7\rm{P}^{\circ}$ levels are $-2.5$\,\% (shown in bold in Table \ref{tab1t}).

We use the resulting CI + all-order wave functions to calculate the electric-dipole reduced matrix elements, with the effective electric-dipole operator $D_{\rm{eff}}$ in the random-phase approximation (RPA). Such effective operator accounts for the core-valence correlations in analogy with the effective Hamiltonian discussed above \cite{DzuFlaKoz98,PorRakKoz99,PorRakKoz99a}.
We also include other correction to the E1 operator beyond RPA;  the core-Brueckner ($\sigma$), two-particle (2P) corrections, structural radiation (SR), and normalization (Norm) corrections \cite{DzuFlaKoz98,PorRakKoz99,PorRakKoz99a}. In most computations of E1 matrix elements, they are omitted, as SR and normalization corrections tend to partially cancel for divalent atoms \cite{DzuFlaSil87}. Reference \cite{Safronova2013} discussed the size of these corrections in Sr, concluding that they cannot be omitted at the 1\,\% level of accuracy.
For Cr, we find that a normalization correction is very large, 11\,\%, which is another consequence of starting from the $V^{N-6}$ potential. All other corrections are small ($< 1$\,\%). The results are listed in Table~\ref{tab2t}. The CI+all-order values include RPA correction, the contributions from the core-Brueckner ($\sigma$), two-particle (2P), structural radiation (SR), and normalization (Norm) corrections are listed separately. Final matrix elements are given in column ``Total''. The transition rates are given in the last column in s$^{-1}$.

 We also studied the effects of saturating the configuration space for all three transitions. In these tests, we compared results of two computations obtained when the odd configurations with weights above $10^{-6}$ and $10^{-7}$ were truncated, corresponding to
44878 relativistic configurations, 13.2 million Slater determinants and 115359 relativistic configurations, 38.6 million Slater determinants, respectively.
 Truncating configurations with small weights is frequently done in CI computations to improve convergence, compute more states, or calculate small corrections. However, our tests show that truncating configurations at the 10$^{-6}$ level may introduce significant differences in cases where mixing is important.
 For $J=4$, truncation of odd configurations with weight smaller than $10^{-6}$ only changed the value of the $\rm{y}^7P_4^{\circ}-a^7S_3$ matrix element by 0.1\,\%, while the $\rm{y}^7\rm{P}_3^{\circ}-a^7\rm{S}_3$ and $\rm{y}^7\rm{P}_2^{\circ}-a^7\rm{S}_3$  matrix elements changed by 2.7\,\% and 4.9\,\%, respectively.
 This sensitivity  is due to mixing with the $^5\rm{P}^{\circ}$ levels for $J=2$ and $J=3$ cases mentioned above. 
%

We confirm the sensitivity of matrix elements to mixing of $^5\rm{P}^{\circ}$ levels by studying  changes in approximate quantum numbers $L$ and $S$ computed as described above. For $J=4$ level, $S= 2.9994$, even if we omit configurations with weights above $10^{-5}$. For $\rm{z}^5\rm{P}_2^{\circ}$, $\rm{y}^7\rm{P}_2^{\circ}$, and $\rm{y}^5\rm{P}_2^{\circ}$ levels, $S$ is equal to 2.033, 2.965, 2.023, respectively, when configurations  with  weights above $10^{-7}$ are included.
These values change to 2.094, 2.859,  2.118, respectively,  when only configurations with  weights above $10^{-6}$ are included. 
We note that the $\rm{y}^7\rm{P}_J^{\circ}-\rm{a}^7\rm{S}_3$ matrix elements are nearly purely nonrelativistic -- i.e. very nearly proportional to $\sqrt{2J+1}$ -- so we expect $L$ and $S$ to be good quantum numbers.

We also calculate transitions from the $\rm{y}^7\rm{P}_J^{\circ}$ levels to other states besides the ground state. This contribution is negligible for $J=4$. The largest contribution to total transition rates for $J=2$ and $J=3$ comes from transitions to the $3d^54s$~$\rm{a}^5\rm{S}_2$ state, with the branching on the order of 1\,\%. We note that mixing strongly affects these small matrix elements, and large CI spaces are needed for branching ratio computations.
  The final total decay rates $\Gamma$ for the $\rm{y}^7\rm{P}_J^{\circ}$ for $J=2,3,4$ are $140\,\times\,10^6$\,s$^{-1}$, $141\,\times\,10^6$\,s$^{-1}$, and $145\,\times\,10^6$\,s$^{-1}$, respectively. 
  We expect slightly worse accuracy for $J=2$ and $J=3$ cases as these are more sensitive to the saturation of the CI space as described above.

\section{Experimental Details}

Our experimental apparatus is based on a pulsed beam of Cr atoms produced using a cryogenic buffer gas beam (CBGB) source \cite{Maxwell2005,Hutzler2011}.  The CBGB source cell dimensions are based on the design outlined in Truppe \textit{et al.} \cite{Truppe2018}.  Atoms are laser ablated from a solid chromium metal sample with a nominally 9\,ns long, 10\,mJ pulse of 532\,nm light.  In our design, the chromium sample is stationary.  The ablation laser focus is repositioned  when the yield of an ablation spot has dropped by roughly half, typically after several thousand ablation pulses. The source is run with a He buffer gas flow rate of 30\,mL/min at standard temperature and pressure.  We orient our experiment by taking  $\hat{z}$ to  be the direction of travel of the atomic beam (roughly parallel to the ground), $\hat{y}$ vertically upward, and $\hat{x}$ parallel to the ground and forming a right-handed coordinate system.

Excitation light is produced using a titanium-doped sapphire laser.  The light is frequency doubled by a lithium triborate crystal in a bow-tie resonator cavity to produce the ultraviolet light resonant with the transitions from the ground state to the  y$^7$P$^{\rm{o}}_J$ states.  An acousto-optic modulator (AOM) allows the laser light to be turned on and off rapidly; we observe a typical turn-off times of 4.5\,ns. In order to minimize the presence of stray light in the experiment, the first-order diffracted beam is directed to the atomic beam vacuum chamber via a 1\,m-long polarization maintaining optical fiber.  The laser beam has a nominal 1/e$^2$ diameter of 15\,mm.  It is retroreflected to intersect the atomic beam twice, propagating in the $\pm\hat{y}$-directions, and polarized linearly in the $\hat{z}$-direction. 

  A hybrid photomultiplier and avalanche photodiode (Picoquant PMA Hybrid-06 \cite{NISTDisclaimer}) images the atomic beam along the $-\hat{x}$ direction.  This detector combines the single photon sensitivity of a photomultiplier (roughly 20\,\% quantum efficiency for the wavelengths of interest here), but has negligible afterpulsing \cite{Wahl2020}, a common systematic effect inherent to standard PMT detectors \cite{Aggarwal2019}.   Because afterpulsing  is not present, the data do not need to be cut to exclude events where multiple fluorescence photons are detected from a single excitation.  However, care must still be taken to ensure the photon count rate is sufficiently low that multiple photons are not incident on the detector within the the detector dead time.  The detector is cooled and typically produces 10 dark counts per second.
A complementary metal-oxide-semiconductor (CMOS) camera viewing along $+\hat{x}$ images fluorescence from the atomic beam on the a$^7$S$_3\rightarrow$ y$^7$P$^{\rm{o}}_J$ transition.  The camera images were used to measure isotope shifts from the observed fluorescence, and  to monitor the ablation yield during decay rate measurements.

\section{Isotope shift measurement}

\begin{figure}[t]
    \centering
    \includegraphics[width=\columnwidth]{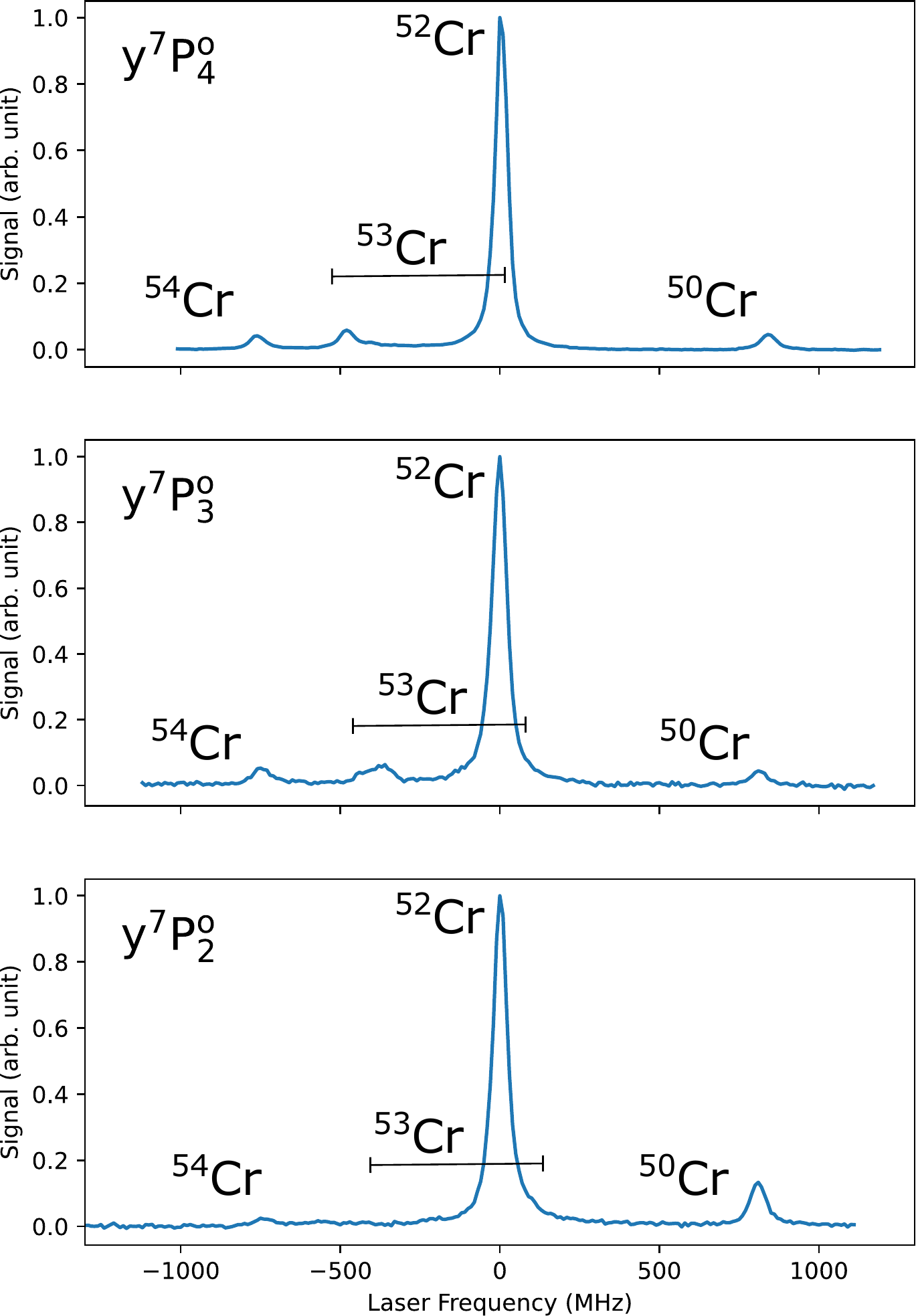}
    \caption{Laser induced fluorescence signal as a function of laser frequency for the a$^7$S$_3 \rightarrow$ y$^7$P$^{\rm{o}}_J$ transitions.  Laser frequency is relative to the fitted $^{52}$Cr center frequency.  Black horizontal bands depict the frequency extent of the $^{53}$Cr hyperfine structure.   }
    \label{fig:spectra}
\end{figure}

The laser is scanned over a roughly 2\,GHz range centered on each transition.  Typical fluorescence spectra for each transition measured by total counts on the CMOS camera  are shown in Fig.\,\ref{fig:spectra}. As is typical with CBGB sources, the yield tends to decrease with a halflife of several thousand ablation pulses, and the observed peak heights in Fig.\,\ref{fig:spectra} do not identically match the natural isotopic abundance of Cr \cite{IsotopeAbundance}.  Systematic asymmetry in the observed lineshapes is mitigated by summing multiple scans of the laser up and down in frequency.  The $^{52}$Cr  and $^{53}$Cr  isotopes are readily identified.  By  collecting fluorescence for multiple sets of 10 ablation pulses with the laser alternately resonant with the $^{50}$Cr (2.4\,\% natural isotopic abundance) and $^{54}$Cr (4.3\,\% natural isotopic abundance) isotopes, we achieve an average signal ratio in good agreement with their natural isotopic abundance ratio. 

The  titanium-doped sapphire laser is referenced to a confocal Fabry-P\`erot cavity  locked to a frequency stabilized helium-neon laser.   The cavity has a nominal 1.5\,GHz free spectral range.  The cavity nonlinearity is measured to be less than 2\,MHz per free spectral range by scanning the laser frequency while applying 115 MHz sidebands using an electro-optic modulator.  

The line centers for the bosonic isotopes are determined by fitting to a Voigt profile with Lorentzian width equal to the measured decay rate $\Gamma$ given in Table \ref{tab:error budget} and detailed in Sec.\,\ref{sec:decay rate}.  Due to the partially resolved nature and low signal to noise ratio of the $^{53}$Cr features, we have not attempted to fit this isotope.  Our measured isotope shifts $\nu_{A-A^\prime} = \nu_A - \nu_{A^\prime}$ are listed in Table \ref{tab:isotope shift}.


\begin{table}[h]
\begin{tabular}{lccc}
\hline\hline
Isotope shift &  y$^7$P$^\circ_2-\rm{a}^7\rm{S}_3$ & y$^7$P$^\circ_3 - \rm{a}^7\rm{S}_3$ & y$^7$P$^\circ_4 -\rm{a}^7\rm{S}_3$\\
\hline
$\nu_{50-52}$   &     807.9(1.4)      &  811.6(1.9)     &   836.8(1.2)       \\
$\nu_{52-54}$   &     738.7(2.8)        &  744.2(2.8)  &  763.3(1.8)\\ 
\hline\hline
\end{tabular} \caption{Measured isotope shifts in MHz for the Cr I y$^7$P$^\circ_J-\rm{a}^7\rm{S}_3$ transitions. Values in parentheses are the combined $1\sigma$ statistical and systematic uncertainty.
}
\label{tab:isotope shift}
\end{table}

\section{Decay Rate measurement}
\label{sec:decay rate}

In this section we detail the measurement and analysis procedures used to determine the decay rates $\Gamma$.  Our measured values of $\Gamma$ for the y$^7$P$^{\rm{o}}_J$ states are presented in Table \ref{tab:error budget} with a list of uncertainty estimates.  

The output of an arbitrary waveform generator in pulse mode is split and sent to both the AOM switching electronics and the sync channel of a multichannel event timer device (Picoquant Multiharp 150 \cite{NISTDisclaimer}).  The output of the photodetector is sent to an input channel of the event timer, and the time of arrival is measured against the sync channel. A histogram of events is collected with 80\,ps time bins by the event timer. 
Timing, computer control of equipment, and data collection are performed using the Labscript Suite  \cite{Starkey2013}.

Each pulse of the cryogenic buffer gas beam of Cr lasts roughly 8\,ms.  At helium flow rates of 30\,mL/min, the temporal distribution of the pulse is observed to be bimodal, with roughly half of the atoms passing through the detection region between 0.8\,ms and 2.8\,ms after ablation.  We therefore typically only collect decay rate data for this fraction of the atom pulse. In the standard measurement configuration, the AOM produces light pulses with 16\,ns full width at half maximum at a pulse rate of 5\,MHz for a duration of 2\,ms (i.e. $10^4$ light pulses per Cr ablation pulse).  The Cr target is ablated at a repetition rate of 15\,Hz.  Data is read out from the multichannel event timer after 300 ablation pulses.  We then perform the same procedure, absent the ablation pulse, to obtain a background scattered light signal.  We define a run of the experiment as 50 iterations of this signal plus background measurement procedure.  A single run lasts roughly 40\,min and achieves about 1\,\% to 2\,\% statistical uncertainty on the decay rate for the state of interest. We performed 6, 12, and 7 runs measuring the decay rate of the $J=2,3,4$ states, respectively.  In addition, several runs were performed on each state with one experimental parameter varied in order to quantify potential systematic errors. 

For the $J=2,3$ states, typical laser power is 10\,mW, and narrow bandpass interference filters are used to monitor the fluorescence decay of the $\lambda\,=\,$ 495\,nm, 494\,nm transitions to a$^5$S$_2$, respectively.  For $J=4$, the transition to  a$^5$S$_2$ is electric dipole forbidden, and we initially attempted to monitor the combined fluorescence decay on the  504\,nm and 509\,nm transitions to the a$^5$D$_{3,4}$ states with another bandpass filter.  However, the y$^7$P$^{\rm{o}}_{4}\rightarrow$a$^5$D$_{3,4}$ signal was observed to be roughly 100 times lower than the
 y$^7$P$^{\rm{o}}_{2,3}\rightarrow$a$^5$S$_2$ decays, in agreement with our calculated transition dipole matrix elements as well as the relative line strengths reported by Wallace and Hinkle \cite{Wallace2009}.  We therefore proceeded  to measure the y$^7$P$^{\rm{o}}_{4}$ decay by monitoring the 358\,nm y$^7$P$^{\rm{o}}_{4}\rightarrow$a$^7$S$_3$ transition using a third interference filter set  with the laser power reduced to 10\,$\mu$W.  This measurement configuration produces similar signal and background count rates as the one used for  y$^7$P$^{\rm{o}}_{2,3}$, but the 358\,nm fluorescence is too weak for useful ablation monitoring with the CMOS camera.


\begin{table}
\begin{tabular}{lddd}
\hline\hline
Parameter ($\times 10^{-6}\,\rm{s}^{-1}$) &  \multicolumn{1}{c}{y$^7$P$^{\circ}_{2}$} & \multicolumn{1}{c}{y$^7$P$^{\circ}_{3}$} & \multicolumn{1}{c}{y$^7$P$^{\circ}_{4}$}\\
\hline
 \vspace{6 pt} Decay Rate $\Gamma$                     & 147.5     & 148.4     & 148.4     \\
 \vspace{6 pt}
 Statistical Uncertainty   &   1.1   &   0.6  &   0.5  \\
 Truncation Error          &   1.0   &   0.5 &   0.4  \\
 Hyperfine Quantum Beats &   0.15  &   0.07 &   0.07 \\
Zeeman Quantum Beats\\
\quad\quad $\mathcal{B}_x$                        &   0.05 &   0.03 &   0.02 \\
\quad\quad $\mathcal{B}_y$                        &   0.3  &   0.26  &  0.5 \\
\quad\quad $\mathcal{B}_z$                        &   0.09 &   0.08  &   0.05 \\
\quad\quad Laser Polarization        &   0.26  &   0.19  &   0.07 \\
Pulse Pileup              &   0.017     &   0.014     &   0.05     \\
 Differential Nonlinearity &   0.017 &   0.018 &   0.018 \\
 Time Calibration          &   0.003     &   0.003     &   0.003     \\ 

 \vspace{6 pt} Systematic Uncertainty, Total         &   1.1   &   0.6  &   0.6  \\
 Total Uncertainty               &   1.6    &   1.0  &   0.8  \\
\hline\hline
\end{tabular} \caption{Measured decay rates $\Gamma$ and $1\sigma$ error budget  for the Cr y$^7$P$^\circ_J$ levels.}
\label{tab:error budget}
\end{table}

\subsubsection{Pulse Pileup}
In order to minimize  missed counts due to coincident events, or ``pulse pileup'', we typically collect data with roughly 1 count per 1000 excitation cycles.  Fluorescence photons are counted using a multihit-capable event timer with a dead time of 650\,ps. The photodetector has a maximum average count rate of 80\,MHz.  Therefore, we correct the raw signal counts $N_i$  in each time bin $i$ to obtain $N_i^\prime$ corrected counts per time bin using
\begin{equation}\label{eq:pulse pileup}
    N_i^\prime = \frac{N_i}{1-\frac{1}{N_{\rm{cycle}}}\sum_{j=i-k_d}^i N_j}
\end{equation}
where $N_{\rm{cylcle}}$ is the total number of excitation cycles, and $k_d = t_{\rm{bin}}\times 80\,$MHz is the number of time bins of width $ t_{\rm{bin}}$ in the detector-limited deadtime \cite{Patting2007}.  We estimate the error $\delta\Gamma$ due to this correction by analyzing the data assuming the detector has no response after the first detection of an excitation cycle.  The difference  is less than $4\times 10^{-4}$ fractional uncertainty $\delta\Gamma/\Gamma$ for our measurements.

\subsubsection{Fit Procedure and Truncation Error}
\begin{figure}
    \centering
    \includegraphics[width=\columnwidth]{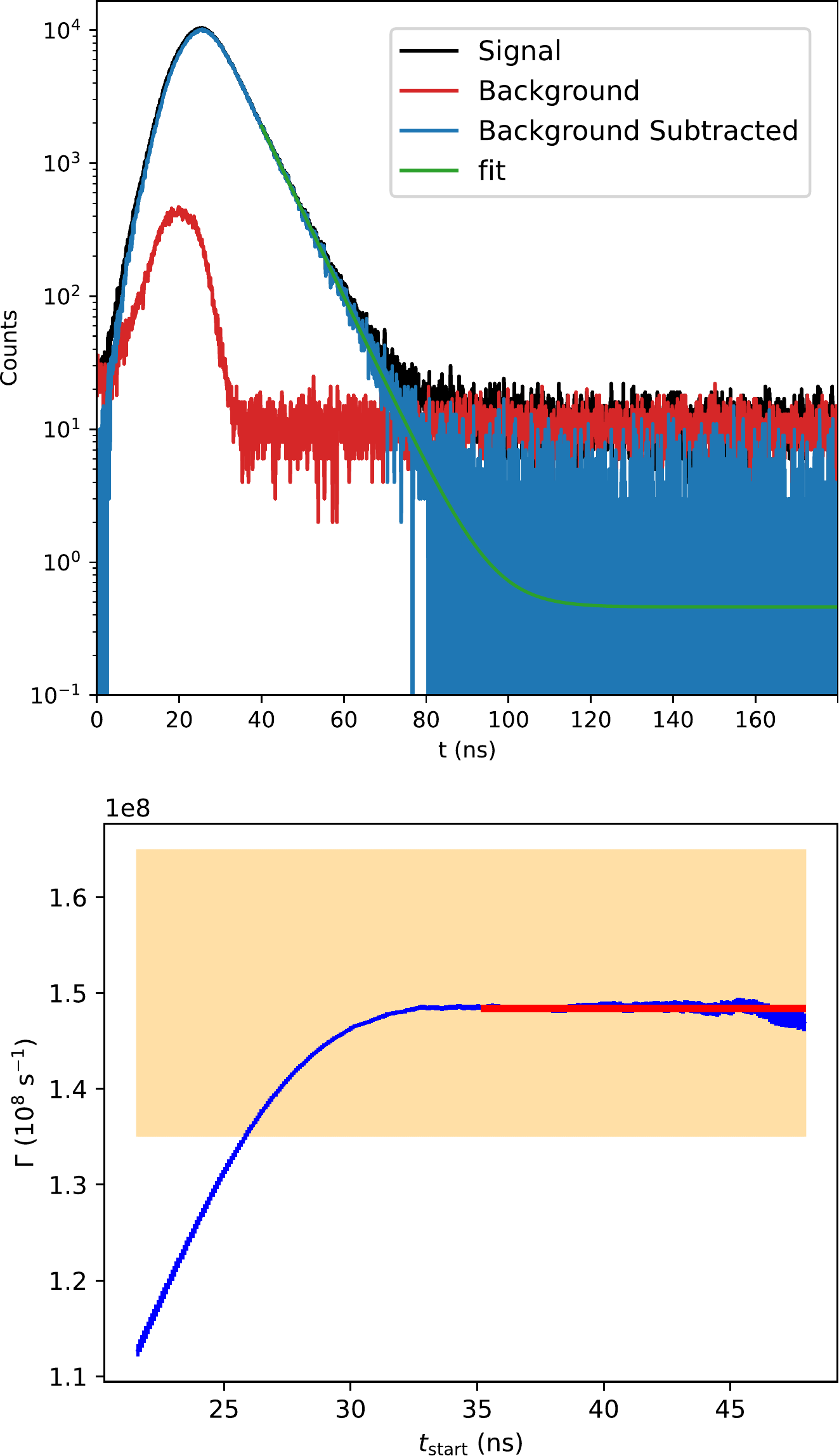}
    \caption{(Top) Histogram of counts as a function of time used to determine decay rate $\Gamma$ for the y$^7$P$^{\rm{o}}_4$ state (358\,nm center wavelength).  The histogram bin width is 80\,ps.  The line labeled ``fit'' is a representative fit to Eq.\,\eqref{eq:single exponential} with $t_{\rm{start}}\,=\,40$\,ns.  (Bottom) Fitted $\Gamma$ as a function of $t_{\rm{start}}$, with error bars indicating the fit uncertainty (blue line). The range of $t_{\rm{start}}$ chosen for our analysis and the standard deviation of fitted $\Gamma$ over this range is indicated by the red line.  The orange band indicates the current consensus range for $\Gamma$ in the NIST atomic spectral database \cite{NIST_ASD}.   }
    \label{fig:Lifetime}
\end{figure}

After correcting for pulse pileup, background counts $N_i^{\rm{bg}\prime}$ are subtracted from the signal counts $N_i^{\rm{sig}\prime}$, and the result is fit to a single decaying exponential plus offset
\begin{equation}\label{eq:single exponential}
    N_i^{\rm{sig}\prime}-N_i^{\rm{bg}\prime} = Ae^{-\Gamma t_i} +c.
\end{equation}
As an example, the pulse-pileup-corrected data measuring the y$^7$P$^{\rm{o}}_{4}$ decay are shown in upper portion of Fig.\,\ref{fig:Lifetime}.

The single largest systematic uncertainty on the measured decay rate is the choice of the start time $t_{\rm{start}}$ of the fit (sometimes called truncation error).   
We vary $t_{\rm{start}}$, setting this parameter to the start of each $t_{\rm{bin}}=80$\,ps time bin over a 12\,ns range which corresponds to between about 30\,\% to 5\,\% of the peak observed counts (Fig.\,\ref{fig:Lifetime} lower portion).  For these late times, the influence of the excitation laser on the decay is negligible.
 In an effort to reduce possible systematic error due to an arbitrary choice of $t_{\rm{start}}$, we assign a value for $\Gamma$ by taking the average fitted $\Gamma$ over this 12\,ns range of start times, weighted by the nonlinear least-squares $1\sigma$ confidence interval when fitting to Eq.\,\eqref{eq:single exponential}.  We assign a statistical uncertainty equal to the median $1\sigma$ nonlinear least-squares fit uncertainty in the  range, and assign a truncation uncertainty equal to the standard deviation of the fitted $\Gamma$ values.

The truncation error is observed to typically be of comparable magnitude to the nonlinear least-squares fit confidence interval for $\Gamma$, and therefore decreases with improved counting statistics.  We  minimize the effects of truncation error by applying the fit to the combined data set of all runs taken under standard conditions.  We also analyzed each data run individually and took a weighted average of the fitted $\Gamma$.  That analysis obtained the same results for each state within the fit confidence intervals.

\subsubsection{Hyperfine Quantum Beats}

The laser pulse has a full width at half maximum of 16\,ns, corresponding to a Fourier limited linewidth of about 100\,MHz.
Decay rate measurements are taken with the laser tuned to the peak in fluorescence which nominally corresponds to resonance with the $^{52}$Cr isotope (84\,\% natural isotopic abundance).   Hyperfine quantum beats are nonexistent for the nuclear spin $I=0$  $^{52}$Cr isotope.  However, hyperfine splitting is present in  $^{53}$Cr ($I=3/2$, 9.5\,\% natural isotopic abundance), and the $^{52}$Cr--$^{53}$Cr isotope shift is sufficiently small that  several $^{53}$Cr lines are not resolved from the  $^{52}$Cr line within the Doppler width of the beam (roughly 20\,MHz) for all three  y$^7$P$^{\rm{o}}_J$ states considered here (Fig.\,\ref{fig:spectra}).  

The hyperfine structure of the y$^7$P$^{\rm{o}}_{2,3}$ states were measured by Becker \textit{et al.} \cite{becker1978}.  The  smallest hyperfine interval is 101\,MHz ($J=2,F=1/2-3/2$), comparable to the Fourier limited laser linewidth.  
While unlikely, we cannot \textit{a prior} rule out the possibility of a small contamination of the signal due to hyperfine quantum beats.

We test for systematic errors due to hyperfine quantum beats by varying the laser detuning by $+20\,$MHz and $-20\,$MHz  from the $^{52}$Cr fluorescence peak.  These detunings correspond to roughly 10 times the root-mean-square error of the laser's cavity transfer lock.   This tests the possibility of different laser frequencies exciting superpositions of $^{53}$Cr hyperfine levels with greater or less efficiency.  In these tests, we detect no statistical difference in the fitted $\Gamma$.  The uncertainties obtained by linear regression are all  $\delta\Gamma/\Gamma < 0.15$\,\%.

Attempts were also made to measure decay rates by exciting  $^{54}$Cr  ($I=0$, 4.3\,\% natural isotopic abundance), which is shifted sufficiently from $^{53}$Cr to rule out potential hyperfine quantum beats.  However, a single run could only provide a statistical uncertainty of roughly $10$\,\%, which was deemed insufficient.

\subsubsection{Zeeman Quantum Beats}

After truncation error, the largest uncertainty in our measurement is the presence of Zeeman quantum beats.    There is no polarizer in the imaging optics, and we did not make an effort to cancel the ambient magnetic field in the detection region, which is measured to be $(\mathcal{B}_x,\mathcal{B}_y,\mathcal{B}_z$)\,$\approx$\,(0.0, 0.2, 0.4)\,G by a three axis Hall probe (1\,G\,$\equiv\,0.1$\,mT).  
The interaction region is surrounded by three pairs of Helmholtz coils which produce a field uniform to 3\,\% over the detection volume.  We test for systematic errors due to Zeeman quantum beats by applying a variable magnetic field along each axis between $\pm 5$\,G.

As expected for our geometry (excitation light polarized along $\hat{z}$ and imaging along $\hat{x}$), when varying $\mathcal{B}_x$ and $\mathcal{B}_z$, we observe no statistical difference in the fitted $\Gamma$.  We estimate error due to these magnetic field components by linear regression assuming an uncertainty of  0.2\,G.  Zeeman quantum beats are definitively observed when $\mathcal{B}_y$ is varied, which produce an inflated truncation error due to the fit amplitude $A$ of Eq.\,\eqref{eq:single exponential} oscillating with fit start time $t_{\rm{start}}$.  For each state, the fitted $\Gamma$ is observed to reach a local extremum around $\mathcal{B}_y \approx 3$\,G; we  therefore empirically fit the fitted $\Gamma$ to a quartic function of $\mathcal{B}_y$ and assume an uncertainty in $\mathcal{B}_y$ equal to its measured value of 0.2\,G to estimate the error in $\Gamma$.

\begin{figure}[t]
    \centering
    \includegraphics[width=\columnwidth]{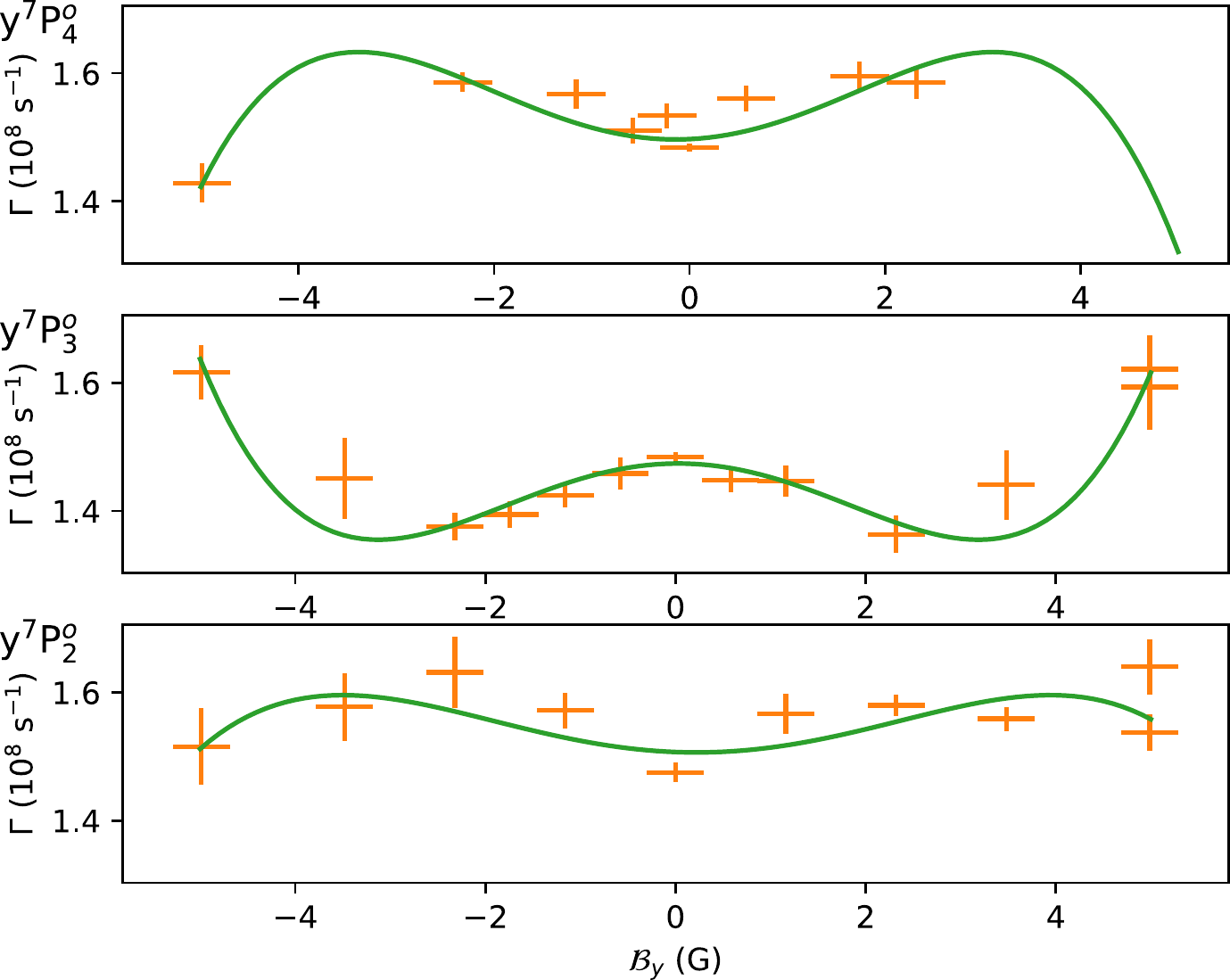}
    \caption{Fitted decay rate $\Gamma$ as a function of applied magnetic field $\mathcal{B}_y$.  The line shows a quartic polynomial determined by a least-squares fit to the data.  Error bars indicate one standard error.}
    \label{fig:By}
\end{figure}

We attempted to model the quantum beats for all three decays.
Because the initial Zeeman phase of the excited state is unknown, it is difficult to predict the quartic fit curves in Fig.~\ref{fig:By} {\it a priori}.
Nevertheless, given that the decay rates and transition wavelengths are roughly equal for all three states, the reduced matrix elements are all roughly the same.
The relative scale of the curves in the panels in Fig.~\ref{fig:By} is then set by the relative Clebsch-Gordan coefficients for the decays that can produce quantum beats.
Our model shows that the amplitude of the beats from y$^7$P$^{\rm{o}}_2$ should be roughly half that of the y$^7$P$^{\rm{o}}_3$ and y$^7$P$^{\rm{o}}_4$, which are roughly expected to be identical.
These expectations in line with the fit curves of Fig.~\ref{fig:By}.

As the ambient magnetic field is not exactly collinear with the applied laser polarization, we further test for Zeeman quantum beats by intentionally rotating the laser polarization to lie along the $\hat{x}$ axis.  In this configuration, we again detect no statistical difference in the fitted $\Gamma$.  The fractional uncertainty obtained by linear regression are all $\delta\Gamma/\Gamma < 0.1$\,\%

\subsubsection{Time Calibration}
The multievent timer has have a differential nonlinearity (that is, non-cumulative uncertainty in the bin width of the time digitizer)  of less than 1\,\%  \cite{Wahl2020}.  This amounts to an uncertainty of 800\,fs, or about $8\times10^{-5}$ fractional uncertainty in the decay rate.  Of secondary concern is  the nonlinearity of the time base of the electronics.  The timing for the experiment is referenced to the internal clock of the TCSPC electronics, which has a specification of $10^{-5}$ stability;  we assign a systematic error of twice this value.  

\section{Comparison of results and conclusion}

\begin{table}[t]
\begin{tabular}{ccccc}
\hline\hline
Method & \multicolumn{3}{c}{Decay Rate $\Gamma (\times 10^6$\,s$^{-1}$)} & Reference \\
\cmidrule(lr){2-4}

&  {y$^7$P$^{\rm{o}}_2$} & \multicolumn{1}{c}{y$^7$P$^{\rm{o}}_3$} & \multicolumn{1}{c}{y$^7$P$^{\rm{o}}_4$} \\
\hline \\ 
 Experiment                     & 147.5(1.6)     & 148.4(1.0)     & 148.4(0.8)  &this work   \\
 \vspace{6 pt} Theory & 140 & 141 & 145 & this work\\
Experiment & 174(17) & 165(17) & 164(16) & \cite{Marek1973}\\
Experiment & 152(7)& 152(7)& 152(7) & \cite{Cooper1997}\\
Experiment & 158(5) &140(4) & 144(4) & \cite{Becker1977}\\

Compilation & 162(16) & 150(15) & 148(15) & \cite{NIST_ASD}\\
\hline\hline
\end{tabular} \caption{Comparison of our measured and calculated decay rates to values from previous works. Values in parentheses are the combined $1\sigma$ statistical and systematic uncertainty.}\label{tab:decay rate comparison}
\end{table}

Here we have presented laser spectroscopy of the y$^7$P$^{\rm{o}}_J$ states of Cr I with a comparison to theory.  We measure the isotope shifts to roughly 0.5\,\% uncertainty.  This appears to be the first reported measurement of these values.  Table \ref{tab:decay rate comparison} compares our measured and calculated decay rates as well as prior reported values.  Our measured decay rates uncertainties are roughly a factor of three better than the next smallest reported uncertainty.
We find excellent agreement, 2\,\% to 5\,\%, between theory and  experiment decay rates, therefore, the differences for the matrix elements are only 1\,\% to 2.5\,\%. We use experimental energies in decay rate calculations. As noted above,  slightly worse accuracy for $J=3$ and $J=2$ is explained by the sensitivity of the mixing with the $^5\rm{P}^{\circ}$ states to the saturation of the CI space as described above. 

We verified that the $J=4$ matrix element value is numerically stable with addition of more configurations. We also tested the accuracy of the effective Hamiltonian by carrying out an additional computation where $H_{\rm{eff}}$ is constructed using second-order perturbation theory, resulting in only 1.7\,\% difference for the $J=4$ matrix element. 
Therefore, such excellent agreement of the $J=4$ value with experiment confirms  our calculations are robust to large (11\,\%) normalization correction, which is important  for further calculations of various properties of atoms and ions with 5 to 6 valence electrons.

These experiments demonstrate the operation of a cryogenic buffer gas beam source, laser system, and detection system suitable for spectroscopic studies of a variety of atomic and molecular species.  We intend to use this apparatus to next perform spectroscopic characterization of the MgF molecule toward laser cooling and trapping.  With minimal modifications, this system could also perform similar studies of more complex molecules such as MgNC, which has been proposed as a candidate for precision measurements of nuclear spin-dependent parity violating physics \cite{norrgard2019,Hao2020}.

\begin{acknowledgments}
The authors thank NIST colleagues Zeeshan Ahmed, Steven Buntin, Tanya Dax, Patrick Egan, James Fedchak, Gerald Fraser,  Nikolai Klimov, Jabez McClelland, William Phillips, Joe Rice, Julia Scherschligt, Ian Spielman,  Stephan Stranick, and Wes Tew for providing equipment used in these experiments.  We thank Chris Billington for software assistance, and thank Mahder Abate and Alessandro Restelli for help with AOM switching electronics. We thank Charles Clark for helpful discussions.  We acknowledge support from the National Institute of Standards and Technology, ONR Grant No. N00014-20-1-2513, US NSF
Grant No. PHY-2012068, and by the Russian Science Foundation under Grant No. 19-12-00157. This research was supported in part through the use of University of Delaware HPC Caviness and DARWIN computing systems: DARWIN - A Resource for Computational and Data-intensive Research at the University of Delaware and in the Delaware Region, Rudolf Eigenmann, Benjamin E. Bagozzi, Arthi Jayaraman, William Totten, and Cathy H. Wu, University of Delaware, 2021, URL: https://udspace.udel.edu/handle/19716/29071.

\end{acknowledgments}

\bibliography{thebib}

\begin{thebibliography}{55}
\bibitem [{NIS()}]{NISTDisclaimer}%
  \BibitemOpen
  \href@noop {} {\bibinfo {title} {{Any mention of commercial products within
  this work is for information only; it does not imply recommendation or
  endorsement by NIST}}}\BibitemShut {NoStop}%
\end{thebibliography}
\end{document}